\documentclass[sigconf]{acmart}

\usepackage{hyperref}

\usepackage[frozencache,cachedir=_minted-main]{minted}
\setminted{fontsize=\small}

\definecolor{myGreen}{rgb}{0.47,0.62,0.11}
\definecolor{myOrange}{rgb}{0.85,0.37,0.008}

\usepackage{enumitem,pifont}
\newlist{todolist}{itemize}{2}
\setlist[todolist]{label=$\square$}
\newcommand{\cmark}{{\color{myGreen}\Large\ding{51}}}%
\newcommand{\xmark}{{\color{myOrange}\Large\ding{55}}}%

\usepackage{tabularray}
\usepackage{cprotect}
\usepackage{numprint}

\AtBeginDocument{%
  }

\acmConference[PAW-ATM '23]{The 6th Annual Parallel Applications Workshop, Alternatives To MPI+X}{November 13, 2023}{Denver, CO}

\begin{document}

\title{Implementing scalable matrix-vector products for the exact diagonalization methods in quantum many-body physics}

\author{Tom Westerhout}
\email{tom.westerhout@ru.nl}
\orcid{0000-0003-0200-2686}
\affiliation{%
  \institution{Institute for Molecules and Materials, Radboud University}
  \streetaddress{Heyendaalseweg 135}
  \city{Nijmegen}
  \country{The Netherlands}
  \postcode{6525AJ}
}


\author{Bradford L. Chamberlain}
\email{bradford.chamberlain@hpe.com}
\orcid{0000-0002-6065-2049}
\affiliation{%
  \institution{Hewlett Packard Enterprise}
  \city{Seattle}
  \country{USA}
}

\begin{abstract}
    Exact diagonalization is a well-established method for simulating small quantum systems. Its applicability is limited by the exponential growth of the so-called Hamiltonian matrix that needs to be diagonalized. Physical symmetries are usually utilized to reduce the matrix dimension, and distributed-memory parallelism is employed to explore larger systems. This paper focuses on the implementation the core distributed algorithms, with a special emphasis on the matrix-vector product operation. Instead of the conventional MPI+X paradigm, Chapel is chosen as the language for these distributed algorithms.

    We provide a comprehensive description of the algorithms and present performance and scalability tests. Our implementation outperforms the state-of-the-art MPI-based solution by a factor of 7--8 on 32 compute nodes or 4096 cores and exhibits very good scaling on up to 256 nodes or 32768 cores. The implementation has 3 times fewer software lines of code than the current state of the art while remaining fully generic.


\end{abstract}

\maketitle

\section{Introduction}

Quantum many-body physics lies at the heart of understanding complex phenomena that emerge from the collective behavior of interacting quantum particles.
In this field, exact diagonalization (ED)~\cite{LinH1990ExactDiagonali} has proven to be a fundamental technique for investigating small-scale systems, serving as a benchmark against which other methods are evaluated.
At its core, ED relies on solving an eigenvalue problem for a sparse Hermitian matrix that is called the \emph{Hamiltonian}~\cite{Griffi2018IntroductionTo}.
The dimension of the matrix grows exponentially with the system size, resulting in an exponential increase in both the required memory and computational resources.

To mitigate the challenges posed by the exponential growth of the Hamiltonian, researchers have employed two key techniques.
First, various symmetries of the system can be exploited to transform the Hamiltonian into a block diagonal form, thereby splitting the problem into multiple ones of reduced dimension~\cite{Sandvi2010ComputationalS}.
Second, distributed-memory parallelism can be leveraged to further increase the system size that can be studied~\cite{Kawamura2017quant,Amaricci2022edipa,Iskako2018ExactDiagonali}.
By combining these techniques, systems of about 48 spin-1/2~\cite{Griffi2018IntroductionTo} particles (hereafter called \emph{spins}) can be studied~\cite{Wietek2018subla,Lauchl2019Kagome}.
This corresponds to a matrix dimension of $2^{48} \approx 3 \times 10^{14}$, but by using symmetries this number can be reduced to about $10^{10}$--$10^{11}$.
Historically, only a handful of research groups have been able to perform such simulations.
The SPINPACK code~\cite{Schule2017Spinpack} is the only open source implementation able to exploit both symmetries and distributed-memory parallelism, but its complexity has limited its adoption.

In an endeavor to democratize the field of exact diagonalization and increase its accessibility, we have introduced the lattice-symmetries package (LS), a user-friendly open-source library designed to facilitate exact diagonalization computations.
Our previous work~\cite{Westerhout2021latti} has already demonstrated that lattice-symmetries achieves state-of-the-art performance on a single node.
Building on this foundation, our focus in this paper is the extension of the lattice-symmetries packages to support distributed-memory parallelism.

For the implementation of distributed algorithms, we chose to use Chapel instead of the more common MPI+X paradigm.
The lead authors of this work initially considered Chapel for this application due to its promise of improved developer productivity.
The language allows for rapid prototyping (compared to MPI) of the algorithms, and performance can then be improved gradually where needed.
Other exact diagonalization packages~\cite{Wietek2018subla,Kawamura2017quant,Schule2017Spinpack} that support distributed-memory parallelism, rely on MPI's collective communication primitives such as \verb|MPI_Alltoallv| and \verb|MPI_Allreduce|,
but we believe that in ED it is actually easier to ensure good work balancing if the algorithms are formulated in terms of asynchronous one-sided communication primitives.
This style of communication is very concisely formulated with tasks in Chapel, because the tasks can access any remote memory that is syntactically in scope without explicitly involving the owner of that memory.

This paper constitutes the first published description of the distributed version of our lattice-symmetries package.
It also includes the first published performance and scalability results relative to what we consider the state of the art in distributed implementations, achieving a $7$--$8\times$ performance improvement on 32 compute nodes or 4096 cores.
It also evaluates the ways in which using the Chapel language has benefited our work, and identifies potential improvements to Chapel that could further benefit lattice-symmetries or other applications in the future.

The rest of the paper is organized as follows: in Sec.~\ref{sec:background} we provide some background information about the problem domain and the Chapel programming language.
Sec.~\ref{sec:related-work} compares lattice-symmetries to the existing packages for exact diagonalization, and Sec.~\ref{sec:software-approach} gives a brief overview of the software used in the implementation.
In Sec.~\ref{sec:algorithms} we delve into the details of our Chapel implementation, covering all the major parallel algorithms, with a particular focus on the matrix-vector product operation, which serves as the cornerstone of all exact diagonalization methods.
Sec.~\ref{sec:performance} discusses the performance and scalability of our implementation.
Finally, in Sec.~\ref{sec:discussion} \& \ref{sec:conclusion} we discuss our experiences in using Chapel instead of MPI+X and mention possible extensions of our work.

\section{Background}\label{sec:background}

\subsection{Exact diagonalization}

Let us start with a quick overview of the key concepts that are relevant to all exact diagonalization methods.
Typically, ED boils down to a (partial) eigenvalue problem for a (potentially) very big Hermitian matrix describing the physical system.
The matrix is called the \emph{Hamiltonian}, and the eigenvalue problem for it is
\begin{equation*}
  H x = \lambda x
\end{equation*}
where $x$ is a vector (called an \emph{eigenvector}), and $\lambda$ is a scalar (called an \emph{eigenvalue}).
$\lambda$ represents the energy of the system when it is in state $x$.
 
If we denote the size of the physical system by $N$, then the Hamiltonian has the dimension $\mathcal{O}(2^N)$.
The origin of this scaling can be understood as follows.
Each spin can be in the basis state $|\uparrow\rangle$ or $|\downarrow\rangle$ or any linear combination of the two: $a |\uparrow\rangle + b |\downarrow\rangle$.
For two spins, the possible states are $|\uparrow\uparrow\rangle, |\uparrow\downarrow\rangle, |\downarrow\uparrow\rangle, |\downarrow\downarrow\rangle$ as well as their linear combinations.
For $N$ spins, we then get $2^N$ different basis states and all their linear combinations, and we can use a vector of $2^N$ numbers to represent any of these linear combinations.
The basis states can be identified with natural numbers or bitstrings of the given length.


For most physical systems of interest, the Hamiltonian $H$ is very sparse: the number of non-zero elements per row or column is typically $\mathcal{O}(N)$.
Additionally, one is usually interested in finding just a few lowest-lying eigenvalues.
For example, the eigenvector corresponding to the smallest eigenvalue describes the system at zero temperature.
The most efficient approaches for finding a few eigenvectors of a large sparse matrix are various Krylov subspace methods (e.g., Lanczos and Arnoldi methods~\cite{Saad2003IterativeMetho}, but also FTLM~\cite{2013StronglyCorrelNumerical}, PRIMME~\cite{Statho2010Primme}).
Even sparse matrix formats such as Compressed Sparse Row (CSR) become impractical, when the system becomes bigger.
For such cases, we switch to a matrix-free representation, where an analytical expression for the Hamiltonian is stored, and the matrix elements are computed on the fly~\cite{Waller2022TrieBasedRank}.
That way, we only store a few vectors of dimension $\mathcal{O}(2^N)$.

\begin{figure}[h]
    \centering
    \includegraphics[width=0.9\linewidth]{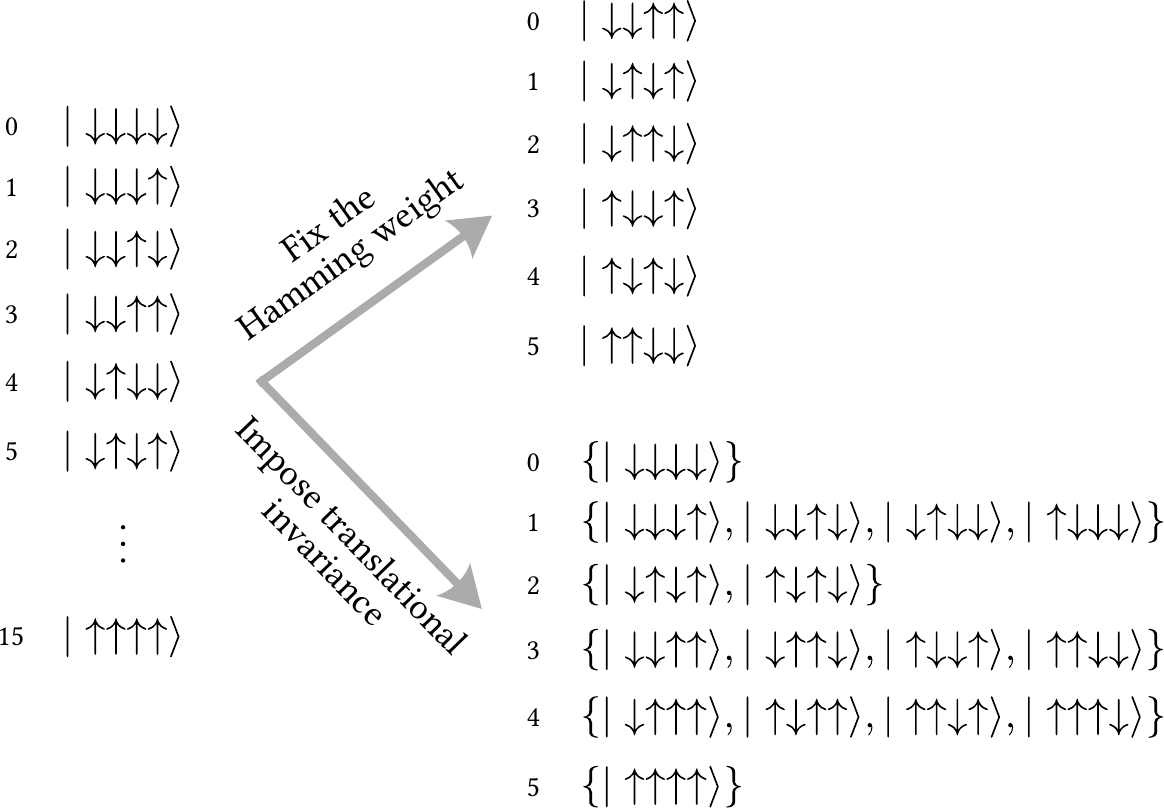}
    \caption{Difference between indices and basis states in the presence of symmetries.}
    \label{fig:symmetries}
\end{figure}

There are no known ways to reduce the asymptotic scaling of the dimension, but we can use physical symmetries to reduce the dimension by a large constant factor.
From the symmetry considerations, we can deduce that the eigenvector of interest lies in a subspace with a smaller dimension and restrict the Hamiltonian to that subspace.

In Fig.~\ref{fig:symmetries}, we give two examples of symmetries for a 4-spin system.
Initially, we had to consider all length-four bitstrings, or 16 basis states.
For each basis state, its index exactly corresponded to the binary representation of the basis state itself (see the left part of Fig.~\ref{fig:symmetries}).

If we know that the system is translationally invariant, the basis states $|\downarrow\downarrow\downarrow\uparrow\rangle$, $|\downarrow\downarrow\uparrow\downarrow\rangle$, $|\downarrow\uparrow\downarrow\downarrow\rangle$, and $|\uparrow\downarrow\downarrow\downarrow\rangle$ all represent the same state.
In practice, we have to store only one coefficient instead of four.
By applying the same trick to all basis states, we reduce the dimension of the space from 16 to 6.
This dimension reduction comes at a cost: the binary representation of our basis states such as $|\downarrow\uparrow\uparrow\downarrow\rangle$ no longer corresponds to their indices, which considerably complicates the implementation as we shall see in Sec.~\ref{sec:algorithms}.

Another example is the so-called $U(1)$ symmetry, which means that we can restrict ourselves to considering bitstrings with a fixed Hamming weight.
This reduces the dimension of our 4-spin system to $\binom{4}{2}=6$ if we fix the Hamming weight to 2 (i.e., two arrows are pointing up).
In practice, the correct Hamming weight is determined from physical considerations.

\begin{table*}[h!]
  \centering
  \begin{tblr}{|c||c|c|c|c|c|c|}
    \hline
    Package & Spins & {Generic \\ Hamiltonians} & {Matrix-free \\ representation} & {Lattice \\ symmetries} & {Distributed-memory \\ parallelism} & {Software lines of code \\ (excluding tests)} \\
	\hline\hline
	lattice-symmetries~\cite{Westerhout2021latti} & \cmark & \cmark & \cmark & \cmark & \cmark & 8500 \\
	SPINPACK~\cite{Schule2017Spinpack} & \cmark & \xmark & \cmark & \cmark & \cmark & 26000 \\
	QuSpin~\cite{Weinbe2017QuspinAPytho,Weinbe2019QuspinAPytho} & {\cmark} & {\cmark} & {\cmark} & {\cmark} & \xmark & 26000 \\
	quantum\_basis~\cite{Wang2023QuantumBasis} & \cmark & \xmark & \xmark & \cmark & \xmark & 12500 \\
	Hydra~\cite{Shackl2022HydraHighPer} & \cmark & \cmark & \cmark & \SetCell[c=2]{c} either one, but not both & & 18000 \\
	libcommute~\cite{Kriven2022LibcommutePyco} & \cmark & \cmark & \cmark & \xmark & \xmark & 4500 \\
	H$\Phi$~\cite{Kawamura2017quant} & \cmark & \cmark & \cmark & \xmark & \cmark & 29000 \\
	Pomerol~\cite{Antipo2015Pomerol} & \xmark & \cmark & \xmark & \xmark & \cmark & 5000 \\
	EDLib~\cite{Iskako2018ExactDiagonali} & \xmark & \xmark & \xmark & \xmark & \cmark & 4000 \\
	EDIpack~\cite{Amaricci2022edipa} & \xmark & \xmark & \xmark & \xmark & \cmark & 11000 \\
	\hline
  \end{tblr}
  \vspace{0.2cm}
  \caption{Feature matrix of various open source exact diagonalization packages. The column ``Spins'' indicates whether the package supports spin-1/2 particles. The column ``Generic Hamiltonians'' indicates whether custom user-defined Hamiltonians are supported by the package and serves as an indication of ease of use and extensibility. Other columns are discussed in the main text.}
  \label{table:feature-matrix}
\end{table*}

\subsection{Chapel}

In this section, we discuss the Chapel programming language that we used to implement distributed algorithms.
Chapel is a programming language that was designed to serve as an alternative to MPI-based parallel programming for large-scale supercomputers.
Chapel supports a global namespace, permitting an expression running on a given \emph{locale}---compute node---to refer to variables stored in the memories of other, remote locales as long as those variables' declarations are lexically visible from the expression's scope.
This permits distributed memory computations to be written without relying on explicit send/receive/put/get calls to transfer data between nodes, as is traditionally done in HPC computations using communication libraries like MPI~\cite{Forum2021mpi} and SHMEM~\cite{Chapman2010intro}.

As a result of this global namespace, Chapel programming tends to look a lot like traditional shared-memory desktop programming.
However, its definition permits programmers to reason about a variable's location, both semantically and computationally.
This supports the ability to optimize for locality and affinity, thereby reducing communication overheads by co-locating computations and the data values they require.
Chapel shares this property with traditional Partitioned Global Address Space~(PGAS) languages like UPC or Fortran 2008~\cite{ElGhazawi2005upc,Fortran2008}.

However, unlike the Single-Program, Multiple Data~(SPMD) programming models used by traditional PGAS languages and MPI, Chapel supports a global view of parallel execution.
This means that a user's Chapel program begins executing as a single task on a single node, which can then create additional tasks, locally or remotely.
This is done using high-level language constructs such as parallel loops or task-parallel abstractions.
These features allow Chapel programs to create new parallel tasks and to migrate them between compute nodes, providing a much more dynamic and general model of parallelism than SPMD-based programming models.

Chapel began as a research project at Cray Inc. within the DARPA High Productivity Computing Systems program in 2003, and began being used in production-grade applications at scale in 2019.
Flagship applications include the Arkouda library for interactive massive-scale data science in Python~\cite{Merrill2019arkou} and the CHAMPS framework supporting aircraft design and analysis using unstructured Computational Fluid Dynamics~\cite{Parent2021DevelopmentOf}.
Chapel is portable and open-source and has been demonstrated to perform similarly to, or better than MPI.
It has also been demonstrated to scale to thousands of nodes and millions of processor cores~\cite{Chambe2023PracticalExamp}.
Recent improvements to the Chapel compiler have enabled its traditional features for locality and parallelism to be applied to GPU programming in a vendor-neutral manner~\cite{Kayrak2023RecentGpuProg}.

\section{Related Work}\label{sec:related-work}

The lattice-symmetries package is not the first open-source code for exact diagonalization, so in Table~\ref{table:feature-matrix}, we provide a feature overview of the most well-known open-source exact diagonalization packages.
For treating the largest systems, it is essential for the package to be memory efficient.
That means, first, using the matrix-free representation of the Hamiltonian because it cuts the memory requirements by a factor $\mathcal{O}(N)$.
Next, one needs to support as many symmetries as possible because the translational invariance alone can reduce the memory requirements by another factor $\mathcal{O}(N)$, and other symmetries reduce the memory requirements even further.
Finally, by using many nodes, one can increase the amount of available memory by a hundred (or even more).

From the feature matrix in Table~\ref{table:feature-matrix}, we see that only SPINPACK implements all of these techniques.
We will thus focus on SPINPACK when comparing the performance of lattice-symmetries to the state of the art.
However, one aspect where SPINPACK could be improved is its ease of use and extensibility.
For instance, with SPINPACK, it is highly non-trivial (and the authors of the current manuscript could not figure it out after studying the code base for some time) to add new interaction types between particles or to compute custom observables.
QuSpin~\cite{Weinbe2017QuspinAPytho,Weinbe2019QuspinAPytho} is a great example of a user-friendly exact diagonalization package and is the most popular choice of all the packages in Table~\ref{table:feature-matrix}.
However, QuSpin's tight integration with Python makes it nearly impossible to add support for distributed-memory parallelism, and its performance is worse than SPINPACK's~\cite{Westerhout2021latti}. We believe that there is a need for a package that achieves state-of-the-art performance and scalability without sacrificing the ease of use, and with the lattice-symmetries package, we try to fill this gap.
To be fair, QuSpin also includes some physics-related features that lattice-symmetries does not.
However, we have not needed these features in our scientific work, and we believe it will be straightforward to add these features to LS while maintaining its architecture and scaling properties when the need arises.
We omit further discussion of these features from this paper due to the scientific complexity required to characterize them for a general HPC audience.

\section{Software Approach}\label{sec:software-approach}

Although the idea of using symmetries to reduce the dimension of the Hamiltonian is not new, our implementation of the distributed algorithms differs drastically from those found in other packages. The tools that we use are also non-standard in either the computational physics or the high-performance computing communities. In this section, we give an overview of the technologies on which the distributed version of the lattice-symmetries package is built.

\paragraph{Multiple languages} We have opted to use multiple programming languages, exploiting the strengths of each one.
Specifically, we have used Haskell for all algorithms that are not performance critical.
These include the compilation of symbolic expressions of the Hamiltonians to low-level kernels, algorithms for building the symmetry groups, and parsing of the input files.
The low-level kernels we implemented in Halide~\cite{Ragan2013Halide} allowed us to easily use SIMD instructions as well as experiment with optimizations such as loop tiling and unrolling.
Chapel was used to implement all shared- and distributed-memory parallel algorithms on top of the Halide-generated kernels.
The interaction between languages was achieved via the C interfaces, large parts of which were generated automatically.
Finally, we also provide a Python interface for users that only require shared-memory parallelism.

\paragraph{Building and deploying} Another non-orthodox choice that we have made is to use the Nix package manager for all steps of the build process.
One of Nix's key strengths lies in its unique approach to package management, utilizing a purely functional and declarative paradigm.
Unlike traditional package managers, Nix builds each package in isolation with all its dependencies, ensuring a consistent and fully reproducible build environment.
Nix allowed us to manage the build steps of different parts of the package, where each step relied on a different build system and external dependencies.
For deploying the application, we have used Apptainer (previously Singularity) containers.
These were built using Nix, too, without relying on Apptainer, thus making Nix the only necessary dependency to build and deploy the code.

In the high-performance computing community, Nix has not been used often, with the usual counter-argument being that not using the system-installed tools might incur a performance overhead at runtime.
In Sec.~\ref{sec:performance}, we show that despite only relying on open-source libraries, we achieve competitive performance relative to the state-of-the-art MPI+X-based solution.

\section{Algorithms}\label{sec:algorithms}

Let us now turn to the description of the distributed algorithms in the lattice-symmetries package.

\subsection{Hashed distribution}

As discussed in Sec.~\ref{sec:background}, in the presence of symmetries, there is a distinction between basis elements and indices.
Although we use integers to represent both, the Hamiltonian matrix is defined in terms of basis elements only.
In order to distribute the basis elements among locales, we introduce a hash function, following the approach of Ref.~\cite{Wietek2018subla}:
\begin{minted}[mathescape=true]{chapel}
proc hash64_01(in x: uint(64)): uint(64) {
  x = (x ^ (x >> 30)) * (0xbf58476d1ce4e5b9:uint(64));
  x = (x ^ (x >> 27)) * (0x94d049bb133111eb:uint(64));
  x = x ^ (x >> 31);
  return x;
}

proc localeIdxOf(basisState : uint(64)) : int {
  return (hash64_01(basisState) % numLocales:uint):int;
}
\end{minted}
In this and all other code listings, we are using Chapel's syntax.
\verb|localeIdxOf| maps a basis element to the corresponding locale index.
By using a hash function that mixes all bits, we achieve very good load balancing.

For input and output (I/O) to disk, we require conversion functions to and from the more common distributions.
We use the block distribution because it is one of the simplest ones, splits the data evenly among the locales, and because it allows for easy parallel reading and writing of files.
In the following, we will discuss our implementation of the conversion functions.

\begin{figure}[h]
    \centering
    \includegraphics[width=\linewidth]{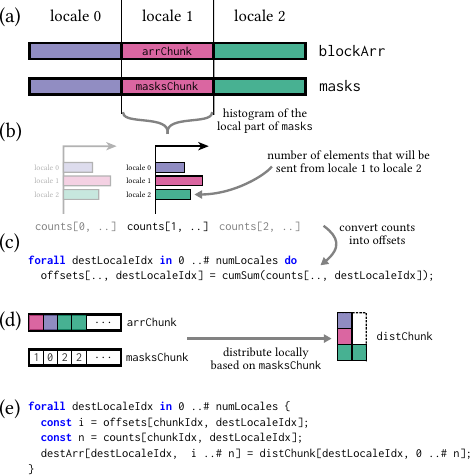}
    \cprotect\caption{Illustration of the algorithm for the conversion of the block-distributed array \verb|blockArr| to a hash-distributed array \verb|destArr|.
    (a) Block distribution of \verb|blockArr| and \verb|masks|.
    (b) Computation of \verb|counts| as histograms.
    (c) Conversion of counts into offsets.
    (d) Locally distributing the chunks according to the values of \verb|masks|.
    (e) Remote put writing to the destination array.}
    \label{fig:algo:arrFromBlockToHashed}
\end{figure}

\paragraph{Block to hashed}

Let us first consider the conversion of a block-distributed array \verb|blockArr| to a ``hash''-distributed array.
To simplify notation, we will describe the algorithm for the case when \verb|blockArr| is one-dimensional.
In the actual implementation, however, we deal deal with both one- and two-dimensional arrays.

As input, we receive \verb|blockArr| and a block-distributed array \verb|masks| that specifies that \verb|blockArr[i]| should be sent to locale at index \verb|masks[i]|.
Importantly, for applications in exact diagonalization, the relative order of elements needs to be preserved.
In other words, if an element \verb|a| appears before an element \verb|b| in \verb|blockArr|, and if both \verb|a| and \verb|b| reside on the same locale in the destination array, \verb|a| should still appear before \verb|b|.

To parallelize the execution, we split the distributed domain into chunks (the total number of chunks will typically equal the product of the number of nodes and the number of cores on each node).
We illustrate this in Fig.~\ref{fig:algo:arrFromBlockToHashed}~(a) for the case of one chunk per node.
For each chunk, we compute the number of elements that will be sent from the current chunk to each locale.
This is accomplished by computing a histogram of the local part of \verb|masks| as shown in Fig.~\ref{fig:algo:arrFromBlockToHashed}~(b).
Then, we convert these counts into offsets by performing a column-wise cumulative sum:

\begin{minted}{chapel}
forall destLocaleIndex in 0 ..# numLocales do
  offsets[.., destLocaleIndex] =
    cumSum(counts[.., destLocaleIndex]);
\end{minted}
These offsets specify the location to where each chunk will write data.
That way, all data transfers can be done in parallel with no need for synchronization.
The next step is locally distributing each \verb|arrChunk|, as shown in Fig.~\ref{fig:algo:arrFromBlockToHashed}~(d).
Finally, parts of \verb|distChunk| are copied to the corresponding locales as shown in Fig.~\ref{fig:algo:arrFromBlockToHashed}~(e).


\begin{figure}[h]
    \centering
    \includegraphics[width=\linewidth]{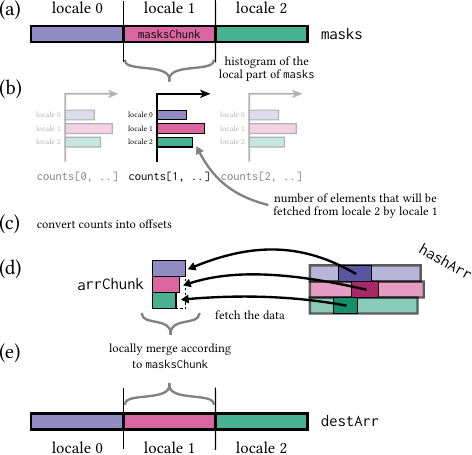}
    \cprotect\caption{Illustration of the algorithm for the conversion of the hash-distributed array \verb|hashArr| to a block-distributed array \verb|blockArr|.
    (a) Block distribution of \verb|masks|.
    (b) Computation of \verb|counts| as histograms.
    (c) Conversion of counts into offsets.
    (d) Fetching the data from \verb|hashArr| into a local \verb|arrChunk|.
    (e) Locally merging the rows of \verb|arrChunk| into the destination array.}
    \label{fig:algo:arrFromHashedToBlock}
\end{figure}

\paragraph{Hashed to block}

In the reverse operation, we are given the \verb|masks| array and a \verb|hashArr| that is distributed among locales according to \verb|masks|.
The goal is to merge the parts into one block-distributed array, but we again need to preserve the ordering of elements.
The algorithm proceeds very similarly to the previous section.
We split the block-distributed domain into chunks, as shown in Fig.~\ref{fig:algo:arrFromHashedToBlock}.
Then, for each chunk, we compute how many elements from which locale need to be copied to it (see Fig.~\ref{fig:algo:arrFromHashedToBlock}~(a)).
These counts are subsequently converted into offsets such that we can use independent remote get operations to fetch the data (see Fig.~\ref{fig:algo:arrFromHashedToBlock}~(b)).
After copying the data (Fig.~\ref{fig:algo:arrFromHashedToBlock}~(c)), we locally merge it and write it to the destination array (Fig.~\ref{fig:algo:arrFromHashedToBlock}~(d)).

\subsection{States enumeration}\label{sec:statesEnumeration}

\begin{figure}[h]
    \centering
    \includegraphics[width=0.9\linewidth]{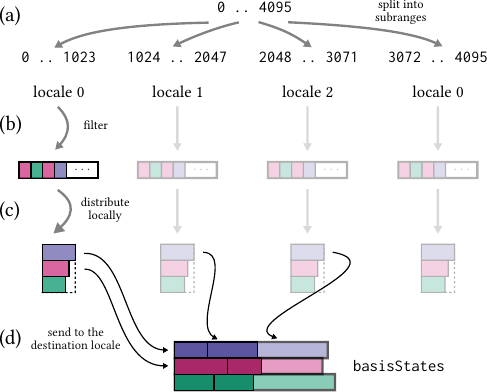}
    \caption{Illustration of the distributed algorithm for enumerating all basis states.}
    \label{fig:algo:enumerateStates}
\end{figure}

Now that we have described the conversion between our internal array distribution based on hashes and the standard block distribution, it is trivial to perform I/O operations and communicate with other packages.
Let us now turn to describe how lists of basis states, such as the ones in Fig.~\ref{fig:symmetries}, are computed.
The goal is to iterate through all bitstrings of a given length, i.e., $2^N$ states where $N$ is the length of the bitstring, and only keep those that satisfy a specific condition.
Discussion of how the condition is implemented is beyond the scope of the current paper, but Refs.~\cite{Wietek2018subla,Sandvi2010ComputationalS} provide some details.

The iteration space can be huge (for instance, for $N=46$, we would need to iterate over $2^{46} \approx 10^{14}$ bitstrings), so it is essential to parallelize the filter operation. 
For this, we split the iteration range into chunks and perform the filter operation on each chunk.
The chunks are distributed among locales in a cyclic fashion for better work balancing since the distribution of the basis states in the original $\{0 \dots 2^N-1\}$ range is highly non-uniform.
The procedure is illustrated in Fig.~\ref{fig:algo:enumerateStates}.
\verb|localeIdxOf| is then applied to each basis state to determine on which locale it should reside.
Afterward, we follow the steps Fig.~\ref{fig:algo:arrFromBlockToHashed}~(b)-(e) to build a hash-distributed array of basis states.
This array is essential for all exact diagonalization algorithms.

\subsection{Matrix-vector product}\label{sec:matrix-vector-product}

Another function that is crucial to any exact diagonalization algorithm is the matrix-vector product:

\begin{equation*}
    y_i = \sum_{j} H_{ij} x_j
\end{equation*}

It is typically also the most time-consuming function.
Let us start with a simple but inefficient implementation because it helps understand the conceptual difference from the standard algorithms that are used for sparse matrices in the CSR format.

\begin{minted}[mathescape=true, escapeinside=@@]{chapel}
// Parallel loop over all locales
coforall destLocale in Locales do on destLocale do
  // Parallel loop over the local part of basisStates
  forall i in basisStates.localSubdomain(destLocale) {
    const @$|\beta\rangle$@ = basisStates[i];
    var acc = 0.0;
    for (@$H_{ij}$@, @$|\alpha\rangle$@) in getRow(@$H$@, @$|\beta\rangle$@) {
      const ref srcLocale = Locales[localeIdxOf(@$|\alpha\rangle$@)];
      // Synchronous remote task spawn
      on srcLocale {
        const j = stateToIndex(basisStates, @$|\alpha\rangle$@); // !!
        acc += @$H_{ij}$@ * x[j];
      }
    }
    y[i] = acc;
  }
\end{minted}

In the code snippet, \verb|basisStates| is the hash-distributed array of basis vectors, \verb|x| and \verb|y| are similarly distributed arrays of real numbers, \verb|getRow| computes all non-zero elements in a row of the Hamiltonian $H$, \verb|stateToIndex| performs a binary search in the local part of \verb|basisStates| to find the index that corresponds to the basis vector $|\alpha\rangle$.
We denote basis vectors with $|\dots\rangle$ and the corresponding indices with Latin letters to help differentiate between the two.

The \verb|stateToIndex| function is the key difference between our representation and the algorithms for CSR matrices or algorithms for matrix-free representations of differentiation operators in various differential equations solvers.
Normally, in the matrix-free representations, one is able to compute the matrix elements corresponding to specific indices $i$, $j$.
Our analytical equation for the Hamiltonian $H$ does not allow that; one is only able to compute the matrix elements corresponding to specific basis states $|\alpha\rangle$, $|\beta\rangle$.
This conceptual difference means that one cannot reuse matrix-vector product implementations from packages such as PETSc~\cite{Balay2023PetscWebPage} and has to start from scratch.

The problem with scaling up the naive implementation is that the information flows both ways: we send $H_{ij}$ and $|\alpha\rangle$ from \verb|destLocale| to \verb|srcLocale|, and then \verb|srcLocale| sends back its contribution to \verb|acc|.
By swapping the order of loops and transposing the Hamiltonian, we can avoid this additional synchronization step:

\begin{minted}[mathescape=true, escapeinside=@@]{chapel}
// Initialize y
y = 0;
// Parallel loop over all locales
coforall srcLocale in Locales do on srcLocale do
  // Parallel loop over the local part of basisStates
  forall j in basisStates.localSubdomain(srcLocale) {
    const @$|\alpha\rangle$@ = basisStates[j];
    forall (@$H_{ij}$@, @$|\beta\rangle$@) in getRow(@$H^T$@, @$|\alpha\rangle$@) {
      const ref destLocale = Locales[localeIdxOf(@$|\beta\rangle$@)];
      const coeff = @$H_{ij}$@ * x[j];
      // Synchronous remote task spawn
      on destLocale {
        const i = stateToIndex(basisStates, @$|\beta\rangle$@);
        // the += needs to be executed atomically
        y[i] += coeff;
      }
    }
  }
\end{minted}

Since the accumulation happens directly in \verb|y[i]| rather than an auxiliary variable \verb|acc|, we initialize \verb|y| to zero.
Note that now we only send $|\beta\rangle$ and \verb|coeff| from \verb|srcLocale| to \verb|destLocale|, but \verb|destLocale| never sends any results back.
This approach maps much better to Chapel's programming model but still has the problem that a remote task is spawned for every matrix element.
The next paragraphs show how we address this challenge.

\paragraph{Computing multiple rows at once.} The first optimization that we apply is replacing the \verb|getRow| function with a \verb|getManyRows| function that computes multiple rows of the Hamiltonian at once.
\verb|getManyRows| receives multiple basis states $[|\alpha_0\rangle, |\alpha_1\rangle, |\alpha_2\rangle, \dots]$ as input, and returns arrays $|\beta_i\rangle$'s and \verb|coeff|$_i$'s.
We then compute the destination locales by applying \verb|localeIdxOf| to every $|\beta_i\rangle$.
We want data transfers to the destination locales to happen in larger chunks, but $|\beta_i\rangle$'s corresponding to a given locale do not necessarily form a contiguous block.
We thus need to first sort $|\beta_i\rangle$'s and \verb|coeff|$_i$'s on their destination locales, but this can be done efficiently in linear time by using an in-place radix sort.

Unfortunately, for every chunk of $|\alpha_i\rangle$'s, we are still spawning a task on each locale.
That means that after just one step, there will be $(\#\mathrm{locales})^2 \cdot \#\mathrm{cores}$ tasks that will be competing for $\#\mathrm{locales} \cdot \#\mathrm{cores}$ cores.
To limit the number of tasks that are spawned, we introduce our second optimization.

\paragraph{Producer-consumer pattern.}
We divide the tasks on each locale into producer and consumer tasks.
The producer tasks execute \verb|getManyRows|, sort the resulting arrays, and send them to their destination locales.
The consumer tasks execute the statements within the inner \verb|on|-clause: they run \verb|stateToIndex| and atomically update the destination vector.
This eliminates the need to spawn remote tasks.
Moreover, we can reuse the buffers for remote transfers, which eliminates the need for re-allocation and pinning.

\begin{figure}
    \centering
    \includegraphics[width=\linewidth]{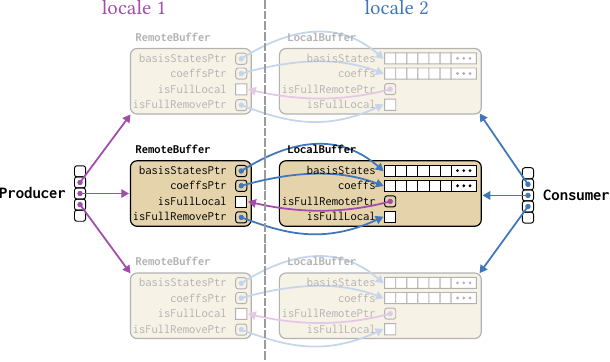}
    \caption{Illustration of the data structures involved in the communication between the producers and consumers.}
    \label{fig:algo:producer-consumer}
\end{figure}

In Fig.~\ref{fig:algo:producer-consumer}, we illustrate how producers and consumers access the data. Each producer keeps pointers to \verb|numLocales| \verb|RemoteBuffer|s. Each \verb|RemoteBuffer| contains pointers to the \verb|basisStates| and \verb|coeffs| arrays on the destination locales. This allows remote data transfers to be done efficiently with remote put operations. Consumers keep pointers to \verb|LocalBuffer|s that store the \verb|basisStates| and \verb|coeffs| arrays. This lets consumers process the data without any communication.

Although elegant, such an architecture means that producers and consumers have to synchronize such that a remote put operation from a producer does not overwrite the data that the consumer has yet to process. We achieve this by using atomics. Each \verb|RemoteBuffer| has a local \verb|atomic(bool)| that specifies whether the buffer is currently full. The \verb|RemoteBuffer| also keeps a pointer to an \verb|atomic(bool)| variable that lives on the destination locale. The producer then does the following:

\begin{minted}[mathescape=true, escapeinside=@@]{chapel}
// wait for isFullLocal to become false
if (remoteBuffer.isFullLocal.compareAndSwap(expected=false,
                                            desired=true)) {
  // generate the data and send it to
  // remoteBuffer.basisStatesPtr and remoteBuffer.coeffsPtr

  // set the atomic on the destination locale
  remoteAtomicWrite(remoteBuffer.isFullRemotePtr, true);
}
\end{minted}
The consumer does the reverse:

\begin{minted}[mathescape=true, escapeinside=@@]{chapel}
// wait for isFullLocal to become true
if (localBuffer.isFullLocal.compareAndSwap(expected=true,
                                           desired=false)) {
  // process the data

  // set the atomic on the source locale
  remoteAtomicWrite(localBuffer.isFullRemotePtr, false);
}
\end{minted}

By always waiting on the local atomic variable, we avoid the overhead of communication, and by first setting the local atomic and then the remote, we ensure consistency and prevent deadlocks. The \verb|remoteAtomicWrite| function can be implemented using network atomics for networks that support it, such as the HPE Cray Aries network. For our use cases, we implemented \verb|remoteAtomicWrite| using active messages but utilizing Chapel's fastOn feature, where the active message is handled directly by the runtime without spawning any tasks.

In conclusion of this section, we would like to remark that the described matrix-vector product algorithm differs drastically from other implementations.
Both the algorithm described in Ref.~\cite{Wietek2018subla} as well as the one used by SPINPACK rely on collective MPI primitives such as \verb|MPI_Alltoallv|.
As a consequence, SPINPACK, for example, does not support overlapping communication with computations.
We believe that it is easier to achieve good work balancing and scaling when using the asynchronous communication model based on the producer-consumer pattern.
The performance of our approach will be analyzed in more detail in the next section.

\section{Performance}\label{sec:performance}

In this section, we analyze the performance of the core algorithms from Sec.~\ref{sec:algorithms}.
We start with the experimental setup.

\paragraph{Hardware} The experiments were performed on the Dutch National Supercomputer called Snellius.
We have used the ``thin'' partition that, at the moment of writing, consists of 504 nodes.
Each node has 2 AMD Rome 7H12 CPUs with 64 cores/socket running at 2.6GHz (in total, 128 cores per node).
The nodes have $16\times 16$ GiB of DDR4 3200MHz memory.
Each node has a single ConnectX-6 HDR100 port.
In other words, the nodes are connected by a 100 Gb/s Infiniband network.

\paragraph{Software} We have used Nix for building the distributed version of the lattice-symmetries package.
That means that we do not rely on any compilers or libraries that are installed globally on the system.
For instance, to build the container that we have used to benchmark the matrix-vector product operation, it is sufficient to issue one command:
\begin{minted}{nix}
    nix build .#distributed.benchmark-matrix-vector-product 
\end{minted}
and the exact same versions of all libraries and compilers will be used as we had in our experiments.
The only system software that we rely on is Apptainer (version 1.1.9-1.el8), that we use for running the containers.
Nonetheless, we list the versions of other packages here for reference:

\begin{itemize}
	\item GCC 12.3.0
	\item LLVM 15.0.7
	\item Halide 15.0.1
	\item GHC 9.6.2
	\item Chapel 1.32.0 pre-release (commit 4585257)
\end{itemize}

For building SPINPACK, we have relied on the system compilers and MPI installation to ensure that we do not negatively affect the performance by using a sub-optimal build of MPI.
Specifically, the code was compiled using GCC 12.3.0 and OpenMPI 4.1.5, built using the same GCC version.

\paragraph{Test Hamiltonians} For measuring the performance of our implementation, we choose to work with closed chains (with periodic boundary conditions) of spin-1/2 particles.
Nearest neighbor spins interact with each other via so-called antiferromagnetic Heisenberg exchange.
We make use of the $U(1)$, spin inversion, translational, and reflection symmetries.
The matrix dimensions for a few systems are presented in Table~\ref{table:hilbert-space}

\begin{table}[h]
  \centering
  \begin{tblr}{|c||r|}
	\hline
    System & Matrix dimension \\
    \hline
    \hline
    40 spins & $\numprint{861725794}    \approx 8.6\cdot 10^8\hspace{0.12cm}$ \\
    42 spins & $\numprint{3204236779}   \approx 3.2\cdot 10^9\hspace{0.12cm}$ \\
    44 spins & $\numprint{11955836258}  \approx 1.2\cdot 10^{10}$ \\
    46 spins & $\numprint{44748176653}  \approx 4.5\cdot 10^{10}$ \\
    48 spins & $\numprint{167959144032} \approx 1.7\cdot 10^{11}$ \\
    \hline
  \end{tblr}
  \vspace{0.2cm}
  \caption{The Hamiltonian matrix dimensions of closed spin-1/2 chains of various sizes.}
  \label{table:hilbert-space}
\end{table}

\subsection{Conversion to and from the hashed distribution}

\begin{figure}[h]
    \centering
    \includegraphics[width=0.9\linewidth]{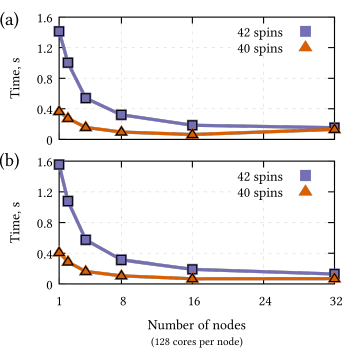}
    \caption{Conversion time between block and hashed distributions. (a) Time of the conversion from block to hashed distribution for 40 (triangles) and 42 spins (squares).
    (b) Time of the conversion from hashed to block distribution for 40 (triangles), and 42 spins (squares).}
    \label{fig:blockHashed}
\end{figure}

The hashed distribution that we are using to store the vectors is an implementation detail of the lattice-symmetries package.
For writing the results to disk as well as communicating with other packages, we need to employ one of the standard distributions.
We chose the block distribution for this purpose, and it is important to verify that the conversion between the block distribution and the hashed distribution does not take longer than the simulation itself.

In Fig.~\ref{fig:blockHashed}, we show the time that it takes to perform the conversion from the block to hashed distribution (a) and back (b).
We use this experiment as a test as well and verify that the roundtrip exactly preserves the vector.
We show the absolute time rather than the performance improvement from using multiple locales.
This is done on purpose because, for this test, we are interested in the absolute time rather than good scaling.

We observe that for more than 4 locales, the operations complete in well under a second, which is negligible compared to other operations.
For example, on 4 locales, a single matrix-vector product for a 40-spin system will take at least 40 seconds.

\subsection{States enumeration}

\begin{figure}[h]
    \centering
    \includegraphics[width=0.9\linewidth]{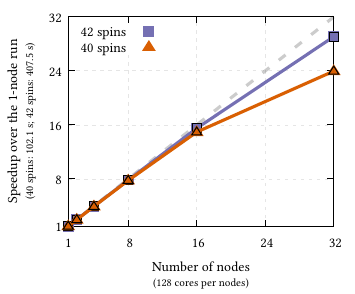}
    \caption{Strong scaling of the basis construction operation. Speedup over the single-locale execution is shown for the 40- and 42-spin systems in circles, triangle, and squares respectively.}
    \label{fig:enumerateStates}
\end{figure}

The next operation that we analyze is constructing the basis.
This involves (see Sec.~\ref{sec:statesEnumeration} for details) of iterating through all possible binary sequences of a given length and filtering those that satisfy some symmetry requirements.

In Fig.~\ref{fig:enumerateStates}, we show strong scaling of the basis construction operation for 40- and 42-spin systems.
We choose these system sizes because they represent the two largest problem sizes we could run on a single node, and science always prefers to run the largest problem sizes possible as they better approximate the results of laboratory measurements.

We observe almost perfect scaling for up to 16 nodes.
For 32 nodes, we see that for the smaller system of 40 spins, the speedup curve starts saturating.
For 42 spins, however, even at 32 nodes, we obtain a very good speedup.
Our hypothesis for the behavior for 40 spins is as follows.

The Hilbert space dimension of the system of 40 spins is $\numprint{861725794}$.
We split the iteration space into chunks such that each core on each node handles around 25 chunks.
After the filtering step (see Fig.~\ref{fig:algo:enumerateStates} for details), each chunk contains around $\frac{861725794}{32 \cdot 128 \cdot 25} \approx 8400$.
Then during the distribution step, we then send $\approx 260$ elements during each remote put.
The average message size is then around 2 KB which is too small to saturate the network bandwidth.
For a system of 42 spins, a similar calculation yields a buffer size of almost 8 KB, which results in better transfer speeds and hence better scaling.

\subsection{Matrix-vector product}

\begin{figure}[h]
    \centering
    \includegraphics[width=\linewidth]{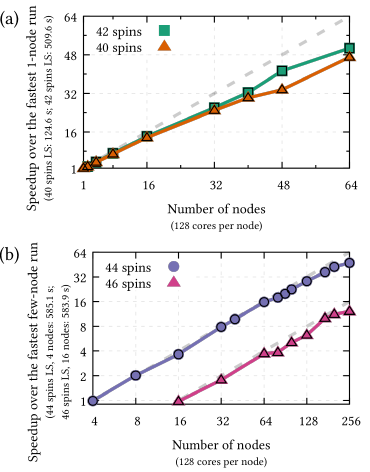}
    \caption{Strong scaling of the matrix-vector product in lattice-symmetries.
    (a) Data points show the speedup over the single-locale execution for 40 and 42 spin systems.
    Dashed gray line indicates perfect scaling.
    (b) Speedup over the 4-node execution for the 44-spin system in circles, and speedup over the 16-node execution for the 46-spin system in squares.}
    \label{fig:scaling-matrixVectorProduct}
\end{figure}

Let us now turn to the scaling analysis of the matrix-vector product operation that was described in Sec.~\ref{sec:matrix-vector-product}.
Basis construction and conversions between block and hashed distributions are executed just a few times during a simulation.
Hence, we care about their performance but do not need them to scale perfectly.
The situation is different for the matrix-vector product because it is often executed hundreds (or in some cases even thousands) of times in exact diagonalization packages.
This section is divided into two parts.
We first discuss the strong scaling of the matrix-vector product operation for both system sizes that still fit into the memory of one node, as well as system sizes that can only be tackled by a distributed-memory implementation.
Then, we compare the performance of our implementation to the SPINPACK package that relies on MPI+X for parallelism.

\paragraph{Scaling} In Fig.~\ref{fig:scaling-matrixVectorProduct}~(a), we show strong scaling of the matrix vector product operation.
The data is normalized by the fastest single-node run with the lattice-symmetries package.
For both 40- and 42-spin systems, we obtain an almost linear speedup for up to 64 nodes (8192 cores).
However, for 42 spins, the speedup we obtain when using 64 nodes is around $51\times$.
The biggest contribution to this slowdown comes from the strict division of tasks into producers and consumers.
For the single-node run, each core spends around 424 seconds generating the matrix elements (\verb|getManyRows|), and around 80 seconds binary searching (i.e., \verb|stateToIndex|) and executing atomic \verb|+=|.
For the 64-node run, each producer (there are $128 - 24$ producers on each node) spends around 8.2 seconds in \verb|getManyRows|.
We see that $\frac{424}{8.2}\cdot \frac{128}{104} \approx 63$, which means that had we used 128 producers instead of 104, we would have gotten an almost ideal $63\times$ speedup.
A very similar calculation holds for the consumers.
We thus conclude that if we allow producers and consumers to steal work from each other to avoid waiting, we expect a performance improvement at 64 nodes that would bring us even closer to the perfect scaling.

Next, we analyze the scaling on system sizes that no longer fit into the memory of a single node.
In Fig.~\ref{fig:scaling-matrixVectorProduct}~(b), in log-log scale, we show strong scaling for 44- and 46-spin systems.
In these experiments, we go to larger node counts because this is where most of the time in the algorithm is spent, so scaling is particularly important.
The data are normalized by the fastest run on the smallest possible number of nodes (4 nodes for 44 spins and 16 nodes for 46 spins).
We observe that our implementation scales well to 256 nodes.
For 44 spins, using 256 nodes, we obtain a $47 \times$ speedup over the 4-node run, and for 46 spins --- a $12\times$ speedup over the 16-node run.

\begin{figure}[h]
    \centering
    \includegraphics[width=\linewidth]{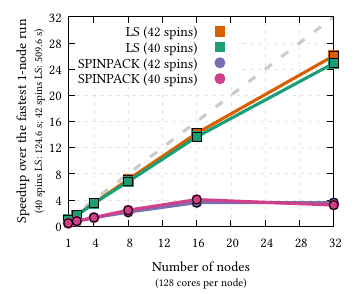}
    \caption{Strong scaling of the matrix-vector product in SPINPACK. We show the speedup over the best single-locale execution for the matrix-vector product for 40 and 42 spins. The data from SPINPACK is shown in circles, and the data from lattice-symmetries (LS) is shown in squares.}
    \label{fig:scaling-spinpack}
\end{figure}

\paragraph{Comparison with MPI-based solutions} We also compare our implementation to the state-of-the-art MPI-based implementation --- SPINPACK.
For this, we perform an analysis similar to Fig.~\ref{fig:scaling-matrixVectorProduct}~(a), except that we now test both lattice-symmetries and SPINPACK.
The SPINPACK package is run in the pure MPI mode, where an MPI process is started for each core.
SPINPACK also supports a hybrid mode in which an MPI rank is run per node, and multithreading is used to drive the cores in parallel, but this did not perform well in our tests.
The results are shown in Fig.~\ref{fig:scaling-spinpack}.
For both 40- and 42-spin systems, the lattice-symmetries package is faster on both one and many nodes.
On one node, lattice-symmetries is $2\times$ faster than SPINPACK, and this factor increases as we increase the number of nodes.
On 32 nodes, lattice-symmetries outperforms SPINPACK by $7$--$8\times$.

\section{Discussion}\label{sec:discussion}

We do not regret our choice of using the Chapel programming language for both shared- and distributed-memory parallelism.
In the following, we list the key features of Chapel that were important for the implementation of the lattice-symmetries package.

\begin{itemize}
  \item We made extensive use of Chapel's global namespace that makes communication between nodes implicit.
        In cases when the Chapel compiler failed to recognize the possibilities of bulk transfers, we made use of the Communication module that provides a portable interface for the low-level remote get and put operations.
  \item Chapel's native support for various distributions has considerably reduced the amount of boilerplate code that we had to write.
        Specifically, we made use of the block and cyclic distributions.
        Multiple times we have considered implementing native support for our hash-distributed arrays but decided against it because it was not user-friendly.
  \item Having a common interface for spawning both local and remote tasks was very helpful, and we were able to express all of our parallel constructs using the \verb|coforall| and \verb|forall| loops.
        We did not use \verb|begin| or \verb|cobegin| partly because they were not necessary and partly because they required manual synchronization (Chapel does have a module implementing futures, but it is not considered stable yet).
\end{itemize}

There are also areas where our implementation can be improved further. First, there is no work stealing taking place between producers and consumers. We believe that adding it would improve performance. It would have been straightforward to implement if Chapel (or, more specifically, the underlying tasking library, Qthreads) had native support for work stealing. There are alternative ways to achieve the desired effect, but we leave their detailed analysis to future work.

Second, in our implementation, we often need to store non-owning references to objects in classes/records. Chapel does not yet support using the \verb|ref| qualifier for class/record attributes, so we rely on raw pointers (\verb|c_ptr| and \verb|c_ptrConst|) instead. This negatively affects both code readability as well as its safety, and it would be interesting to explore possible abstractions on top of raw pointers to restore the safety guarantees that \verb|ref| provides.

\section{Conclusion}\label{sec:conclusion}

In this paper, we have described in detail an implementation of a scalable algorithm for matrix-vector product operation in exact diagonalization applications.
The algorithm utilizes a matrix-free representation for the Hamiltonian and makes use of symmetries to reduce the dimension of the matrix.
By relying on Chapel for parallelization, we achieve good scaling for up to 256 nodes or 32768 cores.
Furthermore, at 32 compute nodes or 4096 cores, we outperform the state-of-the-art implementation based on MPI by a factor of $7$--$8$.
Finally, by combining Chapel with other languages, we obtain approximately a 3-fold reduction in the required number of lines of code while keeping the algorithms generic and the interface user-friendly.

\begin{acks}
The work of T. W. was supported by European Research Council via Synergy Grant 854843---FASTCORR.
We thank SURF (\url{surf.nl}) for the support in using the National Supercomputer Snellius.
\end{acks}

\bibliographystyle{ACM-Reference-Format}
\bibliography{references}

%
%

\end{document}